\documentstyle[prd,aps,psfig,multicol]{revtex}

\newcommand{\gtilde}
 {~ \raisebox{-1ex}{$\stackrel{\textstyle >}{\sim}$} ~}

\begin{document}

\title{Preheating in the Universe Suppressing High Energy Gamma-rays
from Structure Formation
}
\author{Tomonori Totani and Susumu Inoue \\
Division of Theoretical Astrophysics,
National Astronomical Observatory, Mitaka, Tokyo 181-8588, Japan \\
(e-mail: totani@th.nao.ac.jp and inoue@th.nao.ac.jp) \\
\vspace{0.2cm}
{\it Accepted by Astroparticle Physics}
}
\maketitle

\vspace{-4.5cm}
\begin{flushright}
NAOJ-Th-Ap 2001/5
\end{flushright}
\vspace{4.5cm}

\begin{abstract}
Structure formation in the universe can produce high energy gamma-rays from
shock-accelerated electrons, and this process may be the origin of the
extragalactic gamma-ray background (EGRB) as well as a part of 
the unidentified sources detected by
EGRET in the GeV band, if about 5\% of the kinetic energy of the shock
is going into electron acceleration. 
However, we point out that the production of gamma-rays may be severely
suppressed if the collapsing matter has been preheated
by external entropy sources at the time of gravitational collapse, 
as can be inferred from the
luminosity-temperature (LT) relation of galaxy clusters and groups.
We also make a rough estimate of this effect by a simple model,
showing that
the EGRB flux may be suppressed by a factor of about 30.
Hence structure formation is difficult to be the dominant origin of EGRB
if preheating is actually responsible for the observed anomary
in the LT relation.
The detectable number of gamma-ray clusters
is also reduced, but about 5--10 forming
clusters should still be detectable by EGRET all sky, and this number is
similar to that of the steady and high-latitude unidentified sources in the
EGRET catalog. The future GLAST mission should detect 
$10^{2}$--$10^3$ gamma-ray clusters of galaxies even if the
intergalactic medium has been preheated.
\end{abstract}

\begin{multicols}{2}


\section{Introduction}
It is widely believed that the observed structures in the universe
have been produced via gravitational instability. 
Currently the most successful theory of structure formation is the
cold dark matter (CDM) scenario, in which
the structures grow hierarchically from small objects into larger
ones. When an object collapses gravitationally and virializes, the
baryonic matter in the object is heated by shock waves up to the
virial temperature, and particles are expected to be accelerated to high
energy by shock acceleration.
High energy gamma-rays are then expected to be produced
during structure formation, via inverse-Compton
scattering of cosmic microwave background photons by high energy
electrons. Recently, Loeb \& Waxman \cite{lw00} 
speculated that this process can be
the origin of the extragalactic gamma-ray background (EGRB) 
observed at $\sim$ 1--100 GeV \cite{s98}, if about 5\%
of the post-shock thermal energy is going into electron acceleration. 
Totani \& Kitayama (Paper I \cite{tk00}) has shown that, if this is the case, 
a few tens of forming clusters should have already been detected 
by the EGRET experiment which has performed an all sky
survey in the GeV band\cite{h99}, and a part of the unidentified
EGRET sources can be accounted for (see also Waxman \& Loeb \cite{wl00}).

All these analyses are based solely on hierarchical 
structure formation in the
CDM universe, but recent x-ray observations of clusters and groups of
galaxies have shown that the x-ray properties of these objects cannot be
explained by the above simple picture of hierarchical structure
formation alone. 
It is well known that the luminosity-temperature (LT) relation of
clusters and groups is considerably different from what is expected
from the self-similar model predicted by hierarchical structure
formation (e.g., \cite{m98,ae99,hp00}).
The most popular explanation is that the intergalactic
medium has been preheated by external entropy sources such as supernovae
or active galactic nuclei up to a temperature of about 1 keV
(e.g., \cite{k91,bbp99,tn00}).

If this is the case, gamma-ray production from structure formation
should be significantly suppressed, because the external entropy
impedes the gravitational collapse and weakens the shock,
resulting in decreased shock heating and softer spectra of
accelerated particles. It should be noted that the objects on which
preheating has the most significant effect ($M \sim 10^{14} M_\odot$
and $T \sim$ 1 keV) are those which are expected to produce 
most of the EGRB photons.
In this letter we try to make a quantitative estimate of the preheating
effect on the production of gamma-rays from structure formation.

Throughout this paper, we assume a CDM universe with the density
parameter $\Omega_0=0.3$, the cosmological constant
$\Omega_\Lambda=0.7$, the Hubble constant $h = H_0/(100
\mbox{km/s/Mpc})=0.7$, the baryon density parameter $\Omega_B = 0.015
h^{-2}$, and the density fluctuation amplitude $\sigma_8=1$. 
These parameters are favored from various recent cosmological observations.

\section{Effect of Preheating on Structure Formation}
The formulation for calculating the EGRB flux and source counts of
gamma-ray clusters when there is no preheating 
has been given in Paper I. Here we describe the
modification to include the effect of preheating. There are two
important effects of preheating on the production of high energy
gamma-rays from structure formation. The first is that the external
entropy leads to the virialization of collapsing gas at a larger
radius with smaller infalling velocity and smaller kinetic energy of shock
compared with the no-preheating case. The second is that the external
entropy results in a smaller Mach number and a shock 
which is no longer ideally strong,
and hence the energy spectrum of accelerated electrons should be significantly
softer than that without preheating.

It is not easy to predict
realistic density profiles for preheated, collapsing gas
without recourse to detailed numerical simulations,
even in the spherically symmetric, one-dimensional case
(e.g., \cite{kp97,tn00}). In the following discussion,
we do not inquire about the actual distribution of the baryonic matter
within the dark matter halo,
and the relevant physical quantities 
are to be interpreted in a volume-averaged sense.
It should be noted, however, that the production of 
gamma-rays is less sensitive to the gas density profile than, e.g., 
the x-ray luminosity of thermal bremsstrahlung because the target photons 
for inverse-Compton scattering into gamma-rays
are the CMB photons whose density is universal. Thus we believe that
the following simplified treatment is an adequate first approximation
for the estimate of gamma-ray production, although it may be
too simple to calculate accurately the x-ray luminosity and temperature of
clusters of galaxies.

We calculate the virial radius ($r_h$)
and the velocity of collapsing, preheated baryonic gas ($V_h$) 
at virialization
by a simple model naturally extended
from the standard spherical collapse model (e.g., \cite{p80}). 
Let us start from the energy conservation equation of baryonic gas
with mass $M_B = (\Omega_B / \Omega_0)M$ originally embedded in a dark
halo with mass $M$:
\begin{eqnarray}
\frac{1}{2} M_B V^2 - \frac{G M_B M}{r} &+& \frac{M_B a^2}{\Gamma (\Gamma - 1)}
\nonumber \\
&=& - \frac{G M_B M}{2 r_{\rm vir}} + \frac{M_B a_0^2}{\Gamma (\Gamma - 1)} \ ,
\label{eq:energy}
\end{eqnarray}
where $r$ is the characteristic radius of the collapsing gas, and
$V=dr/dt$ is the infalling velocity. We ignore the heating or cooling
during the collapse and assume the collapse proceeds adiabatically, 
and then the effect of preheating is
to add the internal energy $M_B a^2/[\Gamma (\Gamma -1)]$, where
$a = (dP/d\rho_B)^{1/2}$ is the sound 
velocity. The pressure $P$ is related to the density through the entropy
parameter $K$ as $P = K \rho_B^\Gamma$,
and the baryon gas density is $\rho_B = M_B / (4\pi r^3/3)$.
The right hand side of this equation
is the total energy of this system, estimated at maximum expansion
($r = 2 r_{\rm vir}$), where $r_{\rm vir}$ is the virial radius
when there is no preheating. The sound velocity
$a = (dP/d\rho_B)^{1/2}$ is related to that at maximum expansion
$a_0$ as $a  = (r/ 2 r_{\rm vir})^{-1.5 (\Gamma - 1)} a_0$.

The virialization of the system is expected to 
occur at a radius of $r_h$ which is larger than
$r_{\rm vir}$ because of preheating. We estimate this by simply
extending the virial theorem for this system, resulting in the
following equation at virialization:
\begin{eqnarray}
M_B V^2  = \frac{G M_B M }{r} - \frac{3M_B a^2}{\Gamma} \ .
\label{eq:viral}
\end{eqnarray}
Therefore, $V_h$ and $r_h$ are determined by
solving equations \ref{eq:energy} and \ref{eq:viral}. For the case
of $\Gamma = 5/3$, the result is
\begin{eqnarray}
r_h = GM \left[ \frac{GM}{r_{\rm vir}} - 
\frac{9}{5} a_0^2 \right]^{-1} 
\ , 
\end{eqnarray}
and
\begin{eqnarray}
V_h =  \left[ \frac{GM}{r_h} - 
\frac{9}{5} a_0^2 \left( \frac{2 r_{\rm vir}}{r_h} \right)^2
\right]^{\frac{1}{2}} \ .
\end{eqnarray}
The above model is valid only so long as
$r_h < 2 r_{\rm vir}$; haloes with
$r_h > 2 r_{\rm vir}$ cannot collapse from the
point of maximum expansion (turn around) and we assume that 
such haloes have radii $r_h = 2 r_{\rm vir}$
and neither shocks nor gamma-rays are generated.
The original virial radius $r_{\rm vir}$ 
can be calculated by the standard spherical collapse model
for a dark halo of mass $M$ collapsing at $z$, and
then we can calculate the shock energy as the kinetic energy
given to infalling baryonic gas, 
$(3/4) M_B V_h^2$, which is reduced compared with the no-preheating case.

Next we calculate the Mach number of the shock and spectral
index of shock-accelerated particles. In the rest frame of the infall
(i.e., undisturbed) gas, the propagation of the shock can be regarded
as a problem of supersonic piston moving with the velocity of
infall gas, $V_h$, measured in the cluster rest frame. 
The shock velocity $V_s$ (i.e., the upstream gas velocity towards the
shock in the rest frame of the shock front) is given as
(e.g., \cite{l92}):
\begin{equation}
V_s = \frac{\Gamma + 1}{4} V_h + \left[ a^2 + 
\frac{(\Gamma + 1)^2 V_h^2}{16} \right]^\frac{1}{2} \ ,
\end{equation}
and the upstream Mach number is given by ${\cal M} = V_s / a$.
The particle index
is given by $\alpha \equiv - d(\log N_e)/d(\log \gamma_e)
= (r+2)/(r-1)$, where $r = (\Gamma+1)/[(\Gamma - 1)
+ 2/{\cal M}^2]$ is the compression ratio and $\gamma_e$ is the 
electron Lorenz factor.

We must determine an appropriate value of
the entropy parameter, $K$, so that the above model is consistent with
the observed LT relation of clusters and groups. 
We parametrize the entropy parameter as
$K = K_{34, 0} (1+z)^{-1} 10^{34} \rm erg \ cm^2 g^{-5/3}$,
assuming the redshift dependence of $K \propto (1+z)^{-1}$
which is consistent with the observed LT relation\cite{tn00}. 
Following Tozzi \& Norman \cite{tn00}, we take the parameter $K_{34, 0} 
\sim 0.8$ as a fiducial value to be consistent with the LT relation.

\begin{figure}
\begin{center}
  \leavevmode\psfig{file=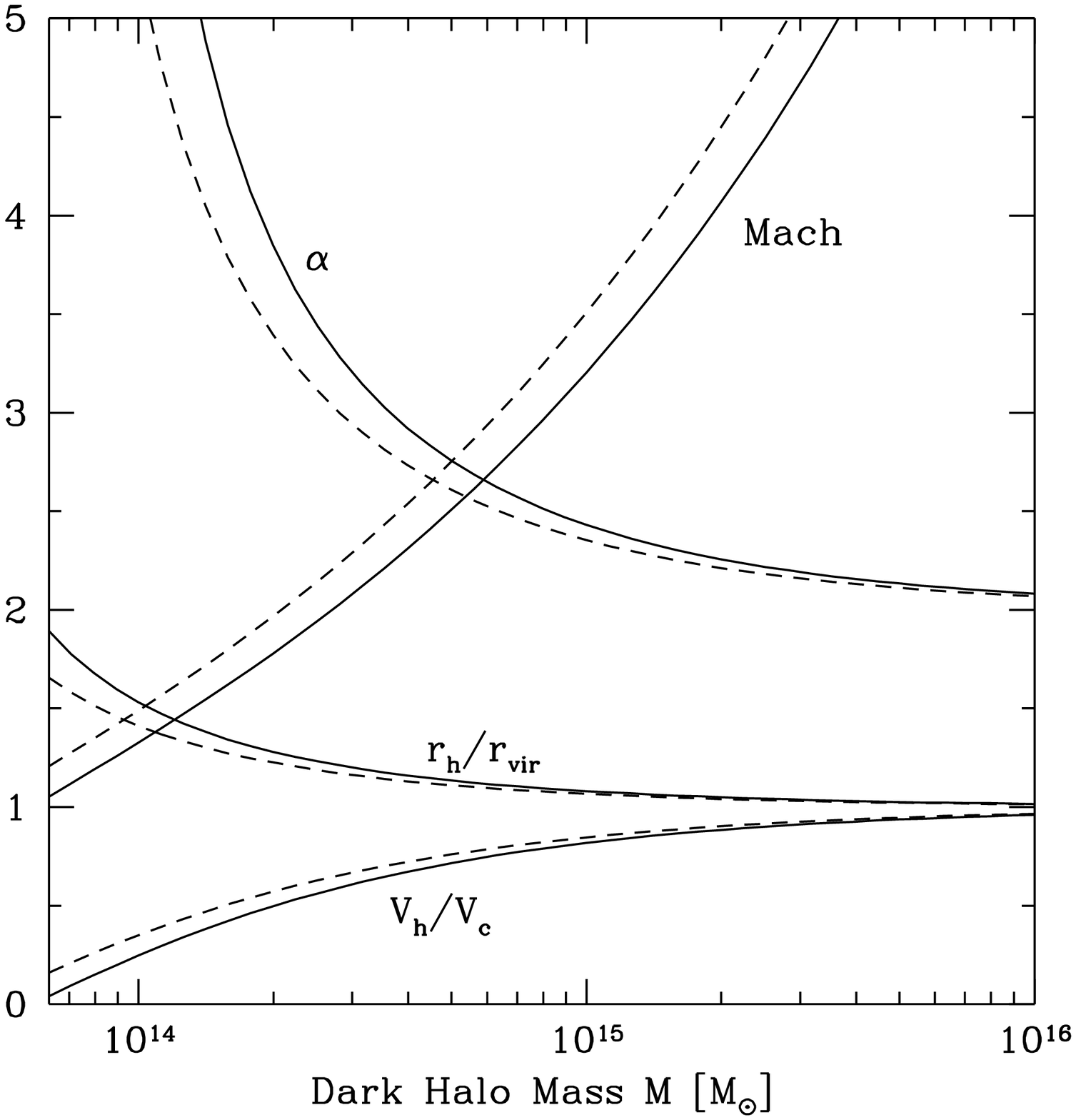,width=12cm}
\end{center}
\caption{The virial radius ($r_h$) and velocity ($V_h$),
Mach number, and particle acceleration index ($\alpha$) of 
gravitationally bound objects under the effect of preheating,
as a function of dark halo mass.
The radius and velocity are given as the ratios to
the original virial radius ($r_{\rm vir}$) and velocity ($V_c$)
in the case of no-preheating. The solid line is for the objects
forming at $z_F = 0$, while the dashed line for $z_F = 1$.
The entropy parameter is assumed
to be $K = K_{34, 0} (1+z)^{-1} 10^{34} \rm erg \ cm^2 g^{-5/3}$
with $K_{34, 0} = 0.8$.}
\label{fig:halo}
\end{figure}

Figure \ref{fig:halo} shows the typical parameters of preheated
haloes as obtained above. It is clear that the objects with $M \sim
10^{14} M_\odot$ is seriously affected by preheating, with the particle
acceleration index much softer than the strong-shock limit of $\alpha = 2$.
Therefore cosmological objects less massive than $\sim 10^{14} M_\odot$
hardly contribute to the high energy gamma-ray background in the GeV band.

As mentioned above, our model may be too simple to calculate the x-ray
luminosity and temperature of preheated haloes, since we have ignored
the density profile within the halo to which the x-ray luminosity is
very sensitive. In spite of this difficulty, however, our model
reproduces the observed LT relation fairly well, as shown below. 
A simple scaling relation for
the x-ray luminosity of thermal bremsstrahlung emission from cluster gas is
$L \propto \rho_B T^{1/2}$ where $T$ is the gas temperature.
It is expected that most of the kinetic energy of collapsing matter
calculated above is eventually converted into thermal energy.
We estimate the temperature so that the final thermal energy is 
the total of this energy from gravitational collapse and
the external energy of the preheating. Then
the temperature of the preheated gas is given by 
$T_h = (V_h/V_c)^2 T_{\rm vir} + \mu m_p K \rho_B^{\Gamma - 1} / k_B$,
where $V_c$ and $T_{\rm vir}$ are respectively the circular velocity and virial
temperature of a halo without preheating predicted
by the spherical collapse model. Here
$\mu$ is the mean molecular weight and $k_B$ is the Boltzman constant.
The X-ray luminosity is then given by $L = (r_{\rm vir}/r_h)^3
(T_h / T_{\rm vir})^{1/2} L_{\rm SS}$, where $L_{\rm SS}$ is the
x-ray luminosity of the self-similar model for non-preheated clusters of
galaxies
which is a function of cluster mass and 
formation redshift. We used the formula of Kitayama \& Suto \cite{ks97}
for $L_{\rm SS}$.

\begin{figure}
\begin{center}
  \leavevmode\psfig{file=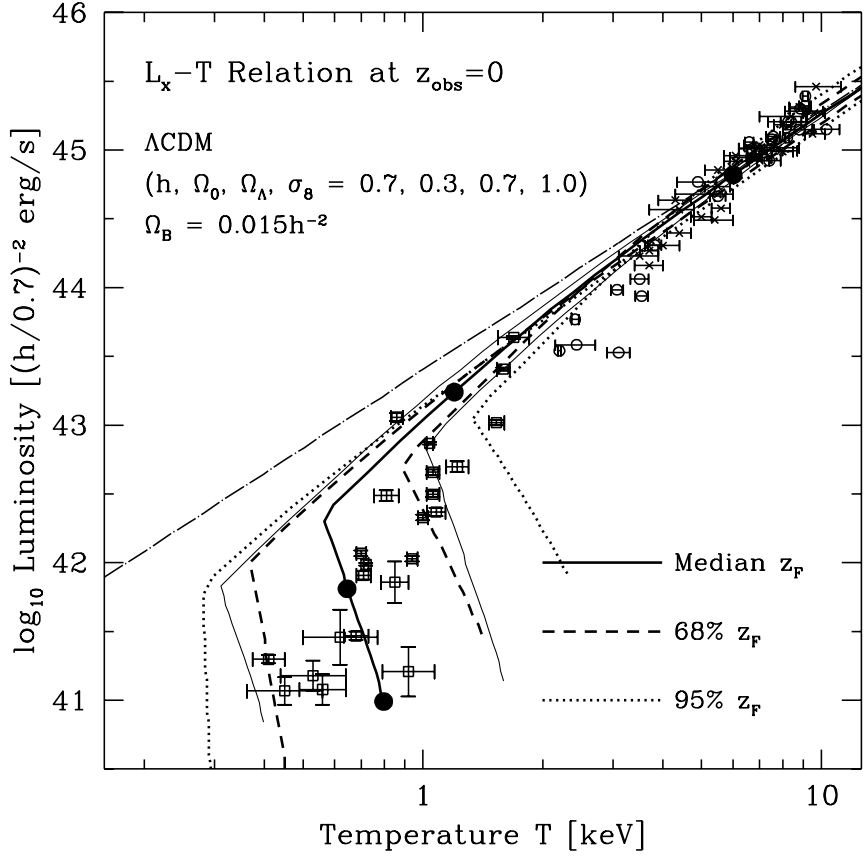,width=12cm}
\end{center}
\caption{The x-ray luminosity-temperature (LT) relation of galaxy clusters
and groups. The thick-solid line is the model prediction for objects observed
at $z_{\rm obs}=0$, using the median of the Lacey \& Cole
\protect\cite{lc93}
distribution function
for the formation redshift ($z_F)$. The entropy parameter is assumed
to be $K = K_{34, 0} (1+z)^{-1} 10^{34} \rm erg \ cm^2 g^{-5/3}$
with $K_{34, 0} = 0.8$. The solid circles are indicating
the grids corresponding to the cluster masses of $10^{12},  10^{13}, 
10^{14}$, and $10^{15} M_\odot$.
The dashed and dotted lines show the dispersion of the LT relation
due to that of $z_F$,
in which $z_F$ is included by a probability of 68\% (1 $\sigma$) and
95\% (2 $\sigma$), respectively. The two thin-solid lines are the same
as the thick-solid line, but for different values of the entropy 
parameter, $K_{34, 0}$ = 0.4 and 1.6.
The dot-dashed line is the prediction of the self-similar model
of x-ray luminosity of clusters without preheating, assuming 
$z_F = z_{\rm obs} = 0$. The data points are from Markevitch 
(cross, \protect\cite{m98}), Arnaud \& Evrard (open circle, 
\protect\cite{ae99}), and 
Helsdon \& Ponman (open square, \protect\cite{hp00}).
}
\label{fig:LT}
\end{figure}

Then we can calculate the luminosity and temperature of a preheated halo
with mass $M$ and collapsing at redshift $z_F$. Figure \ref{fig:LT}
shows the LT relation predicted for cosmological objects observed at
$z_{\rm obs} =0$, compared with observations. Objects should have
various $z_F$ even for the same mass, and here we utilize the
distribution of $z_F$ as a function of mass and $z_{\rm obs}$
derived by Lacey \& Cole \cite{lc93}. 
The thick solid line shows the LT relation
when the median of the Lacey-Cole $z_F$ distribution is applied, using
the standard value of the entropy
parameter: $K_{34, 0} = 0.8$.
The behavior of the model LT curve changes abruptly at 
$(L, T) = (10^{42.3}{\rm erg \ s^{-1}}, 0.6 \ {\rm keV})$, and
this corresponds to the point at which $r_h$ becomes equal to
the maximum expansion radius, $2 r_{\rm vir}$, and hence haloes
cannot gravitationally collapse.
The dashed and dotted lines show the dispersion of the LT relation
corresponding to that of $z_F$, encompassing 68\% (1 $\sigma$) and
95\% (2 $\sigma$) of the probability distribution of $z_F$, respectively. 
The thin solid lines are for median $z_F$, but with different values
of $K_{34, 0}$ = 0.4 and 1.6.
Figure \ref{fig:LT} shows that our simple model is in reasonable agreement
with the data when the same entropy parameter as Tozzi \& Norman
\cite{tn00} is used.

Now we can calculate the EGRB flux and source counts of gamma-ray
clusters by the formulation given in Paper I,
simply replacing the shock energy and particle acceleration index
by those obtained above. 
In the following we will apply the above model to calculate the 
EGRB flux and spectrum, and the expected counts of gamma-ray clusters.

\section{Extragalactic Gamma-Ray Background}
The efficiency of energy injection from the kinetic energy of
infalling gas into nonthermal electrons by the shock acceleration
acceleration is parametrized by $\xi_e$, and
we assume 5\% injection, i.e., $\xi_e = 0.05$ following Loeb \& Waxman
\cite{lw00} and Paper I. It is widely accepted that 
low-energy cosmic rays are accelerated in supernova remnants, and
its energy injection efficiency is about 1--10\% for ions. 
On the other hand, injection into electrons is still highly
uncertain, both observationally and theoretically. As is well known,
the energy flux of cosmic-ray electrons observed above the Earth's atmosphere
is about 100 times lower than
that of cosmic-ray protons, and it may suggest that energy injection
is considerably lower for electrons than for ions. 
However, the local cosmic-ray flux ratio
does not necessarily reflect that at the production
site, because of different energy loss and propagation processes. 
Hence, we take
$\xi_e = 0.05$ as a maximally possible value, although it could be rather
optimistic.

\begin{figure}
\begin{center}
  \leavevmode\psfig{file=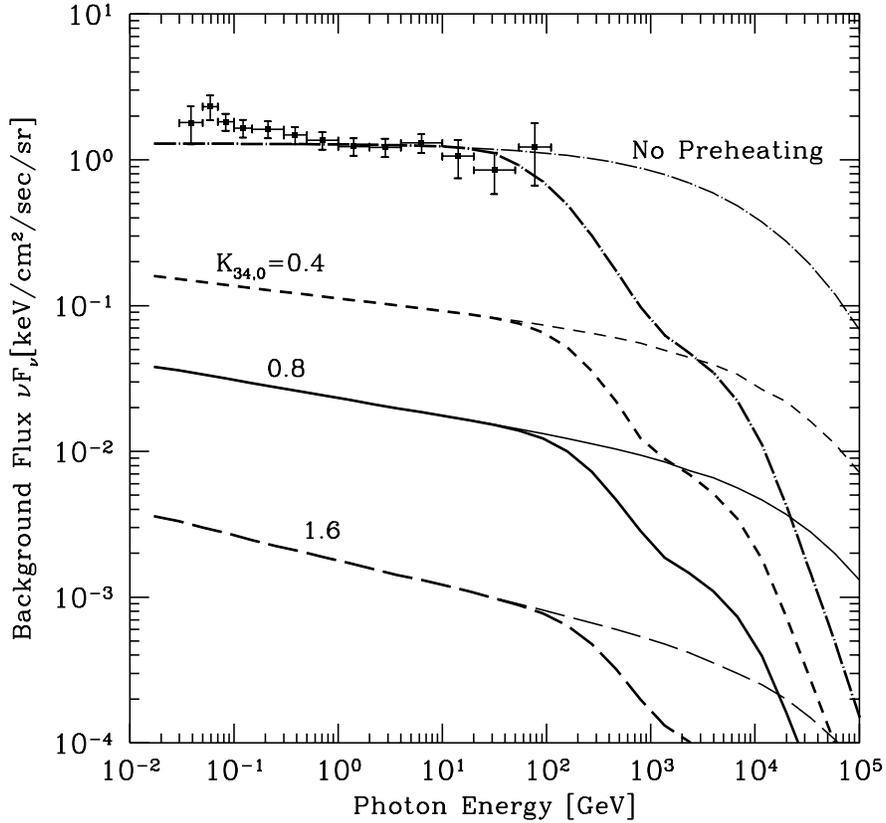,width=12cm}
\end{center}
\caption{The spectrum of the cosmic gamma-ray background (EGRB)
in the GeV band expected from structure formation. 
The short-dashed, solid, and long-dashed lines are for the 
entropy parameters of $K_{34, 0}$ = 0.4, 0.8, and 1.6, where
$K = K_{34, 0} (1+z)^{-1} 10^{34} \rm erg \ cm^2 g^{-5/3}$.
($K_{34, 0}=0.8$ best fits to the observed LT relation of galaxy clusters.)
The dot-dashed line is the result of Paper I without the effect of
preheating. All thick lines take into account the absorption of
gamma-rays in the intergalactic field using the opacity presented
in Totani (2000)\protect\cite{t00}
while the thin lines do not. (Reproduction of gamma-rays 
is not taken into account in either case, see Paper I). The observed data
are from Sreekumar et al. (1998)\protect\cite{s98}.
}
\label{fig:CGB}
\end{figure}

Figure \ref{fig:CGB} shows our calculation of the EGRB flux and spectrum
based on the model described above. We have used
a value of the magnetic field parameter, $\xi_B = 10^{-3}$,
which is the ratio of magnetic energy to the total
gravitational energy given to baryonic matter. Magnetic fields 
of $\xi_B \sim 10^{-3}$ are sometimes observed in intracluster matter,
and the EGRB spectrum extends up to $\sim 100$ GeV if $\xi_B \gtilde 10^{-5}$
(Paper I). Note that this parameter determines only the
maximum photon energy of the EGRB spectrum, and the EGRB flux is rather 
insensitive to this uncertain parameter. It can be seen
that the EGRB flux is severely decreased by a factor of about 30
for preheating of $K_{34, 0} \sim 0.8$, 
which fits best to the observed
LT relation. Even if we use a relatively small
value of $K_{34, 0} = 0.4$, the EGRB flux is about one order of 
magnitude smaller than that observed. Therefore, if the
intergalactic medium is actually preheated, the structure formation
is very unlikely to be the origin of EGRB. It should be noted that
we have already assumed a relatively large efficiency of shock energy
injection into electron acceleration, $\xi_e = 0.05$, and hence 
we cannot take the option of increasing this parameter to save this hypothesis.

\section{Forming Gamma-Ray Clusters of Galaxies}
\begin{figure}
\begin{center}
  \leavevmode\psfig{file=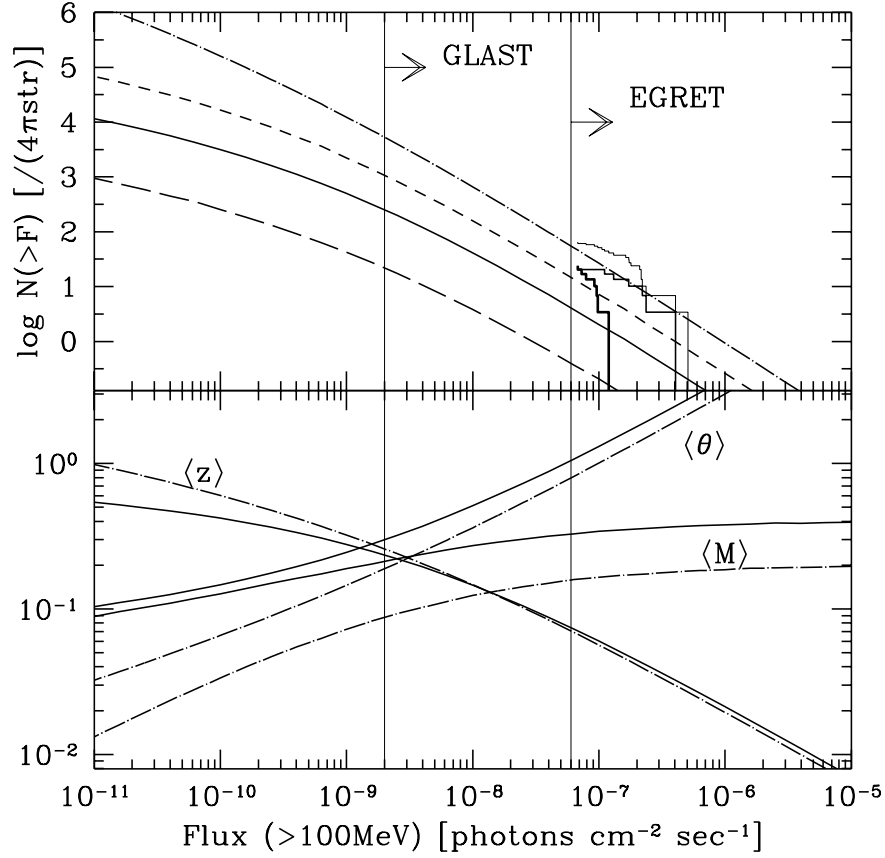,width=12cm}
\end{center}
\caption{
The upper panel: the cumulative flux distribution of
gamma-ray emitting clusters of galaxies. The four curves are for 
different values of the entropy parameter, $K = K_{34, 0} (1+z)^{-1} 
10^{34} \ \rm erg \ cm^2 \ g^{-5/3}$, with $K_{34, 0}$ = 0 (no preheating,
dot-dashed), 0.4 (short-dashed), 0.8 (solid, best fits to
the LT relation), and 1.6 (long-dashed). The observed distribution of the
unidentified EGRET sources with $|b|>45^\circ$ is shown by the three
solid lines, corresponding to all unidentified sources, `em' sources
(see Paper I), and steady unidentified sources defined by Gehrels et al.
(2000)\protect\cite{g00}, 
with the order of the line thickness from thinner to thicker.
The sensitivity limits of the EGRET and GLAST experiments are shown
in the figure. 
The lower panel: the mean redshift, cluster mass (in units of
$10^{16}M_\odot$), 
and angular radius (in degree) of gamma-ray clusters brighter
than a given flux are shown. The line markings are the same as
the upper panel.}
\label{fig:counts}
\end{figure}

The top panel of Fig. \ref{fig:counts} shows the calculation of the source
counts for gamma-rays from structure formation. Compared with the 
case of no preheating, the source counts are decreased by a factor of
10 above the EGRET sensitivity limit.
See the bottom panel of the figure for the mean values of physical
quantities of gamma-ray emitting objects (mass, redshift, and angular
radius) brighter than a given flux.
The mean spectral index for gamma-ray clusters detectable above
100 MeV is rather insensitive to the flux, increasing as
$\alpha$ = 2.20 to 2.26 from the sensitivity of the EGRET to
that of GLAST, because only objects
with sufficiently hard spectrum can be observed at GeV energies.

For the canonical value of the entropy parameter, $K_{34, 0} = 0.8$,
the number of gamma-ray clusters detectable by EGRET is about 5 in all
sky. If we take a relatively small value of $K_{34, 0} = 0.4$ 
to account for the model uncertainty, the number is increased to
14. These numbers are smaller than those reported by Paper I because of
the preheating effect, but it is interesting to note that the number of
steady (i.e., unlikely to be AGNs) 
unidentified sources of the EGRET catalog presented in
Gehrels et al. \cite{g00}
is 7 for $|b| > 45^\circ$ ($\sim 24 \pm 9$ in all
sky), which is not very different from our result. A significant part
of these high-latitude, steady unidentified EGRET sources could be
explained by dynamically forming gamma-ray clusters.

\section{Conclusion}
We have shown that the preheating of intergalactic medium, which may
have occurred as indicated from the observed x-ray properties of clusters and
groups of galaxies, significantly suppresses the production of high-energy
gamma-rays from structure formation. If preheating is actually
responsible for the steepening in the LT relation, 
structure formation cannot be the dominant origin of
EGRB. The number of discrete sources detectable by the EGRET
is also decreased by preheating, but 5--10 gamma-ray clusters could still be 
observable all sky, which may constitute a part of the unidentified sources. 
Even if the preheating effect is profound,
the future GLAST mission may detect about 100--1000 gamma-ray clusters,
and it may be used to probe the preheating processes in the
intergalactic medium as well as the dynamical processes of structure 
formation.

The authors would like to thank N. Gehrels and D. Macomb
for providing their data for steady unidentified EGRET sources.
We would also like to thank helpful discussions with E. Waxman.
TT has partially been supported by the Grant-in-Aid for the
Scientific Research Fund (No. 12047233) of the Ministry of Education, Science,
and Culture of Japan.


\end{multicols}

\newpage

\end{document}